\pdfoutput=1

\documentclass[11pt]{article}

\usepackage[utf8]{inputenc}
\usepackage[T1]{fontenc}
\usepackage{amsmath,amssymb,amsfonts}
\usepackage{graphicx}
\usepackage{booktabs}
\usepackage{natbib}
\usepackage{geometry}
\usepackage{hyperref}
\usepackage{float}
\usepackage{caption}
\usepackage{subcaption}
\usepackage{threeparttable}
\usepackage{multirow}
\usepackage{enumitem}
\usepackage{xcolor}

\hypersetup{
    colorlinks=true,
    linkcolor=blue!60!black,
    citecolor=blue!60!black,
    urlcolor=blue!60!black,
    pdfauthor={Mingdeng Du},
    pdftitle={Tiered Super-Moore's Law: Price Evolution, Production Frontiers, and Market Competition in LLM Inference Services}
}

\title{%
\textbf{Tiered Super-Moore's Law: Price Evolution, Production Frontiers,\\
and Market Competition in Large Language Model Inference Services}%
}

\author{
Mingdeng Du\\
\textit{Wuhan University}\\
\texttt{martindu@whu.edu.cn}
}

\date{March 2026}

\begin{document}

\maketitle

\begin{abstract}
\noindent
This paper provides the first systematic economic analysis of token pricing in the large language model (LLM) inference market. Assembling a novel dataset integrating OpenRouter API data (318 models), Epoch AI records (3{,}237 models), and 62 cross-validated milestone observations spanning 2020--2026, we document an approximately 600-fold decline in token prices and propose the ``Tiered Super-Moore'' hypothesis. Economy-tier models exhibit a price half-life of 1.10 years and mid-tier models 1.55 years---both significantly faster than Moore's Law's two-year benchmark---while flagship models display near-zero exponential fit ($R^2 = 0.031$) due to a reasoning premium averaging 31.5 times non-reasoning prices. A Chow structural break test identifies May 2024 as the critical market inflection point ($F = 5.74$, $p = 0.005$), marking a transition from technology-driven to competition-driven price acceleration. Cost decomposition reveals that total factor productivity residuals account for approximately 103.7\% of cost reduction, with GPU hardware contributing only $-0.9$\%, confirming that software and architectural innovation---not hardware advances---drive the decline. Data Envelopment Analysis shows a Malmquist Productivity Index peaking at 4.11 during 2024Q1--Q4, with technological frontier shift ($TC = 4.13$) as the dominant driver. Training cost--inference pricing elasticity is $\beta = 0.432$, and the 63-fold training cost gap between U.S.\ and Chinese firms is statistically attributable to architectural innovation (\$/FLOP difference insignificant, $p = 0.228$) rather than factor price differentials. Market concentration declined sharply, with HHI falling from 4{,}558 to 2{,}086 over three years. These findings establish token economics as a distinct subfield of digital goods pricing and carry implications for competition policy, AI accessibility, and international technology governance.


\end{abstract}

\section{Introduction}\label{sec:intro}

The commercialization of large language model (LLM) inference services has created a new category of digital commodity: the \textit{token}. Since OpenAI launched the GPT-3 API in June 2020 at \$60 per million input tokens, the market for LLM inference has expanded to encompass dozens of providers, hundreds of models, and pricing that spans four orders of magnitude. By early 2026, economy-tier models such as Gemini 2.0 Flash offer comparable or superior quality at \$0.10 per million tokens---a 600-fold price decline in under six years.

Despite the economic significance of this market---AI inference revenue is projected to exceed training revenue by a substantial margin---rigorous economic analysis remains scarce. The literature on AI economics has focused predominantly on labor displacement \citep{acemoglu2018race, acemoglu2019automation, autor2015}, aggregate productivity effects \citep{brynjolfsson2014second}, and the cost of training frontier models \citep{sevilla2022compute, epoch2024training}. Token pricing as a research object has received almost no systematic treatment. This is a consequential gap: token prices determine who can access AI capabilities, how quickly AI diffuses through the economy, and how the surplus from AI progress is distributed between producers and consumers.

This paper fills the gap by providing the first comprehensive empirical study of LLM token pricing. Our contributions are fourfold:

\begin{enumerate}[leftmargin=*, itemsep=2pt]
\item \textbf{The ``Tiered Super-Moore'' hypothesis.} We document that token prices decline faster than Moore's Law, but only in economy and mid-tier segments (half-lives of 1.10 and 1.55 years versus the two-year Moore benchmark). Flagship models defy exponential decay due to reasoning premiums averaging 31.5$\times$.

\item \textbf{Structural break identification.} We identify May 2024 as a statistically significant structural break ($F = 5.74$, $p = 0.005$), marking the transition from technology-driven to competition-driven price acceleration, catalyzed by the DeepSeek price war and the GPT-4o-mini launch.

\item \textbf{Production frontier analysis.} Using DEA and Malmquist indices, we show that the inference production frontier shifted outward at an extraordinary rate (MPI $= 4.11$ in 2024Q1--Q4), driven almost entirely by technological change rather than efficiency catch-up.

\item \textbf{Training--inference cost nexus.} We estimate a training cost--inference pricing elasticity of $\beta = 0.432$ and demonstrate that the 63-fold U.S.--China training cost gap originates from architectural innovation (MoE), not factor price differentials (\$/FLOP difference insignificant at $p = 0.228$).
\end{enumerate}

The remainder of the paper proceeds as follows. Section~\ref{sec:theory} develops the theoretical framework. Section~\ref{sec:data} describes data sources and methodology. Section~\ref{sec:results} presents empirical results. Section~\ref{sec:robust} reports robustness checks. Section~\ref{sec:discussion} discusses implications. Section~\ref{sec:conclusion} concludes.

\section{Theoretical Framework}\label{sec:theory}

\subsection{The Token as a Digital Good}\label{subsec:token_good}

A token is the atomic unit of LLM inference commerce. Unlike traditional digital goods \citep{shapiro1999information, varian2000buying}, tokens possess a unique combination of properties that necessitate a dedicated analytical framework.

First, tokens are \textit{non-storable and non-transferable}. A purchased token is consumed during inference and cannot be resold, eliminating secondary markets and arbitrage---a sharp departure from digital content goods (e-books, software licenses) that can be shared or resold.

Second, tokens exhibit \textit{extreme vertical quality differentiation}. A token processed by GPT-4o-mini (\$0.15/M) differs qualitatively from one processed by o3 (\$10/M) or Claude Opus (\$15/M). This quality ladder spans two orders of magnitude in price and corresponds to measurable differences in benchmark performance, reasoning depth, and output reliability.

Third, the marginal cost of token production displays \textit{strong economies of scale with capacity constraints}. Once a model is trained (a sunk cost ranging from \$1M to \$400M), the marginal cost of inference is determined by GPU compute, memory bandwidth, and energy---costs that decline with batch size and utilization rate but face hard constraints from hardware availability.

These properties place tokens at the intersection of several economic literatures: information goods pricing \citep{varian1995pricing, varian1997versioning}, experience goods with quality uncertainty \citep{nelson1970, akerlof1970}, and infrastructure services with high fixed costs and low marginal costs \citep{rifkin2014zero}.

\subsection{The ``Tiered Super-Moore'' Hypothesis}\label{subsec:hypothesis}

Moore's Law \citep{moore1965, moore1975, schaller1997moores} established that semiconductor transistor density doubles approximately every two years, implying a price half-life of roughly two years for equivalent computing performance. \citet{nordhaus2007two} extended this framework to show that computing costs have declined even faster than Moore's original prediction over certain periods.

We hypothesize that LLM token prices follow a \textit{tiered} version of this pattern:

\begin{description}[leftmargin=2em, labelindent=1em]
\item[H1 (Economy/Mid Tier):] Economy and mid-tier token prices decline with half-lives significantly shorter than two years, exhibiting a ``Super-Moore'' effect driven by architectural innovation, inference optimization, and competitive pressure.

\item[H2 (Flagship Tier):] Flagship token prices do not follow exponential decay. Instead, they are governed by innovation-driven reasoning premiums and versioning strategies \citep{varian1997versioning}, resulting in wave-like dynamics rather than monotonic decline.

\item[H3 (Structural Break):] The token pricing regime exhibits a structural break corresponding to the onset of cross-border price competition, after which the decline rate accelerates.

\item[H4 (Production Frontier):] The LLM inference production frontier shifts outward faster than implied by hardware improvements alone, with total factor productivity (TFP) as the dominant driver.
\end{description}

\subsection{Production Frontier Framework}\label{subsec:frontier}

Following \citet{solow1957technical}, we model the LLM inference production function as:
\begin{equation}\label{eq:production}
Q = A(t) \cdot f(K, L, E)
\end{equation}
where $Q$ is inference output (tokens per second at a given quality level), $K$ is capital (GPU hardware), $L$ is labor (engineering effort for model and inference optimization), $E$ is energy, and $A(t)$ is a technology index capturing architectural innovations (MoE, Flash Attention, speculative decoding, quantization).

The key insight is that $A(t)$ has grown far faster than improvements in $K$ alone. The MoE architecture, for instance, reduces active parameters from 671B to 37B during inference \citep{deepseek2025v3}, delivering an 18-fold reduction in per-token compute requirements without proportionate quality loss. This architectural innovation operates independently of hardware cost trends and represents the core mechanism behind the Super-Moore effect.

\section{Data and Methodology}\label{sec:data}

\subsection{Data Sources}\label{subsec:data_sources}

We assemble what is, to our knowledge, the most comprehensive dataset on LLM token pricing to date, integrating three complementary sources.

\textit{OpenRouter API (cross-section).} We collected real-time pricing data for 318 models on March 28, 2026, via the OpenRouter API. Each record includes model identifier, vendor, input price (\$/M tokens), output price (\$/M tokens), and context window. OpenRouter aggregates models from all major providers (OpenAI, Anthropic, Google, DeepSeek, Meta, xAI, Alibaba/Qwen, Mistral, and others), with prices standardized to \$/M tokens.

\textit{Epoch AI (panel).} From Epoch AI's Notable AI Models database, we obtained records for 3{,}237 models covering 2020--2026, including training cost (USD), training compute (FLOP), parameter count, hardware configuration, frontier status, open-weight status, and region (U.S./EU vs.\ China).

\textit{Cross-validated milestones (time series).} We manually compiled and cross-validated 62 milestone pricing records using vendor pricing pages, OpenRouter quotes, and industry reports. The sample covers OpenAI (22 records, GPT-3 through GPT-5.4-pro), Anthropic (14, Claude 1 through Claude Opus 4.6), Google (8), DeepSeek (5), xAI (4), Meta (5), and Alibaba/Qwen (4).

\subsection{Price Tier Classification}\label{subsec:tiers}

We classify models into three tiers based on market positioning and price level:
\begin{itemize}[itemsep=2pt]
\item \textbf{Flagship} ($>$\$5/M input): GPT-4, Claude 3 Opus, o1/o3 series, GPT-5-pro.
\item \textbf{Mid} (\$0.5--\$5/M): GPT-4o, Claude Sonnet, Gemini Pro series.
\item \textbf{Economy} ($<$\$0.5/M): GPT-4o-mini, DeepSeek-V3, Gemini Flash series.
\end{itemize}

\subsection{Econometric Methods}\label{subsec:methods}

\textit{Exponential decay.} For each tier $j$, we estimate:
\begin{equation}\label{eq:decay}
P_j(t) = P_{0,j} \cdot e^{-\lambda_j t}
\end{equation}
where $P_j(t)$ is the input token price, $t$ is time in years since the first observation, $\lambda_j$ is the decay constant, and the half-life is $t_{1/2} = \ln 2 / \lambda_j$. A half-life shorter than two years supports the Super-Moore hypothesis.

\textit{Chow structural break test.} We test for structural breaks in the pooled log-price series by computing the Chow F-statistic at each candidate break date within the sample window, selecting the date with the highest F-statistic.

\textit{Data Envelopment Analysis (DEA).} We employ the CCR (constant returns to scale) model with blended price as input and quality score (0--100) as output to estimate efficiency scores $\theta_i$ for each model. The Malmquist Productivity Index (MPI $= EC \times TC$) decomposes cross-period productivity change into efficiency catch-up ($EC$) and technological frontier shift ($TC$).

\textit{OLS and panel regression.} For the training cost--inference pricing nexus, we estimate:
\begin{equation}\label{eq:training}
\ln(P_{\text{inference},i}) = \alpha + \beta \cdot \ln(C_{\text{training},i}) + \gamma \cdot X_i + \varepsilon_i
\end{equation}
where $C_{\text{training}}$ is training cost and $X$ includes controls for parameter count and region. HC3 heteroskedasticity-robust standard errors are reported alongside OLS standard errors.

\section{Empirical Results}\label{sec:results}

\subsection{Price Trends: A 600-Fold Historical Decline}\label{subsec:trends}

\begin{figure}[htbp]
\centering
\includegraphics[width=0.95\textwidth]{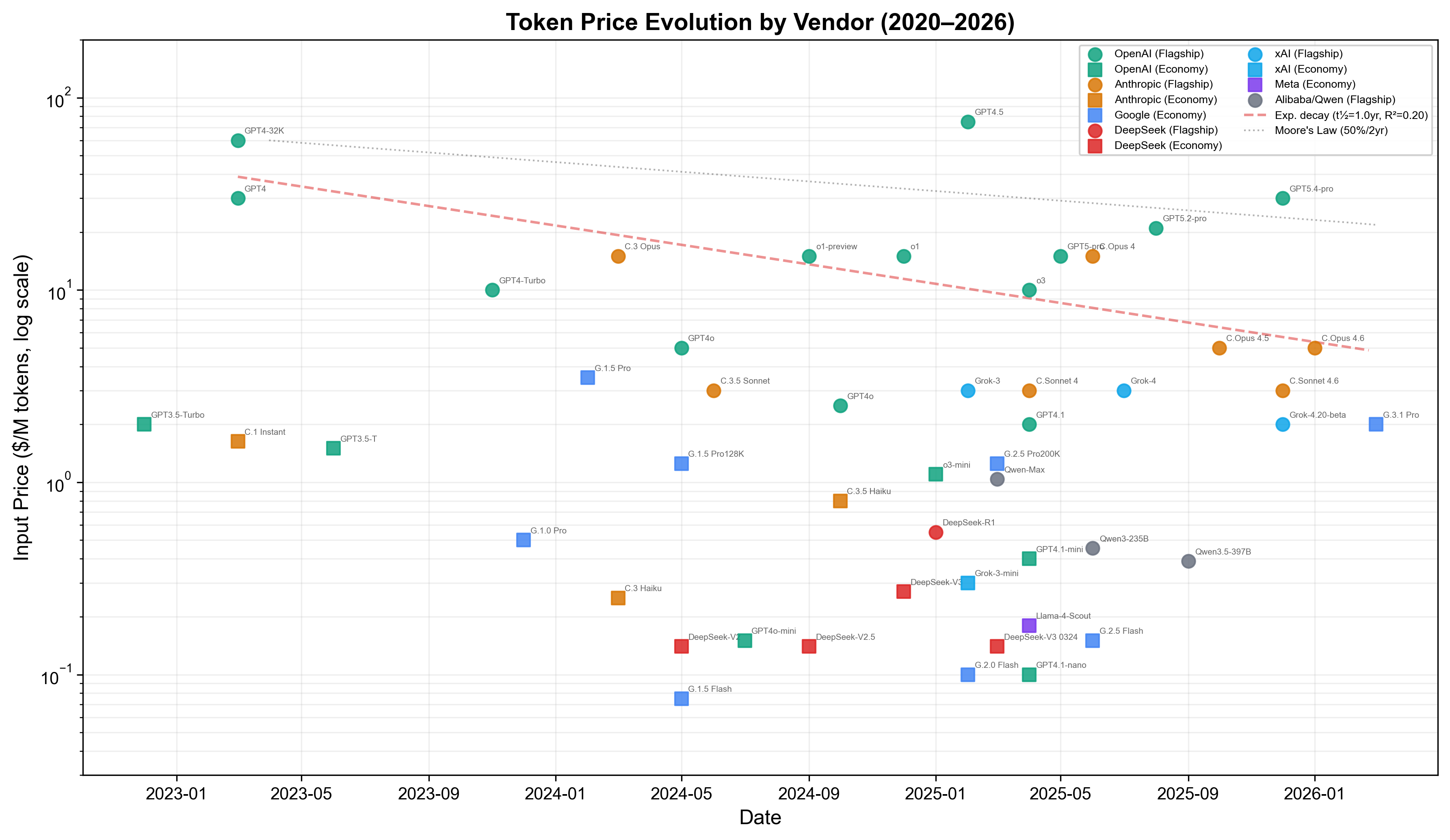}
\caption{Token Price Evolution by Tier (2020--2026). Each point represents a milestone pricing record. The three tiers---flagship, mid, and economy---exhibit sharply different trajectories. The vertical dashed line marks the May 2024 structural break.}
\label{fig:price_evolution}
\end{figure}

Table~\ref{tab:summary} presents summary statistics of the token pricing dataset. The most salient feature is the extraordinary range: input prices span from \$0.01/M (Liquid lfm models) to \$150/M (o1-pro), a 15{,}000-fold range. The median price declined from approximately \$30/M in 2023Q1 to under \$0.50/M by 2026Q1. Figure~\ref{fig:price_evolution} plots the complete price trajectory by tier.

The headline decline from GPT-3 (\$60/M, June 2020) to Gemini 2.0 Flash (\$0.10/M, February 2025) represents a 600-fold reduction. Crucially, this is not merely a nominal price decline---quality-adjusted prices have fallen even faster, as the \$0.10/M Gemini 2.0 Flash outperforms the \$30/M GPT-4 on most benchmarks.

\subsection{Tiered Exponential Decay}\label{subsec:tiered}

\begin{figure}[htbp]
\centering
\includegraphics[width=0.95\textwidth]{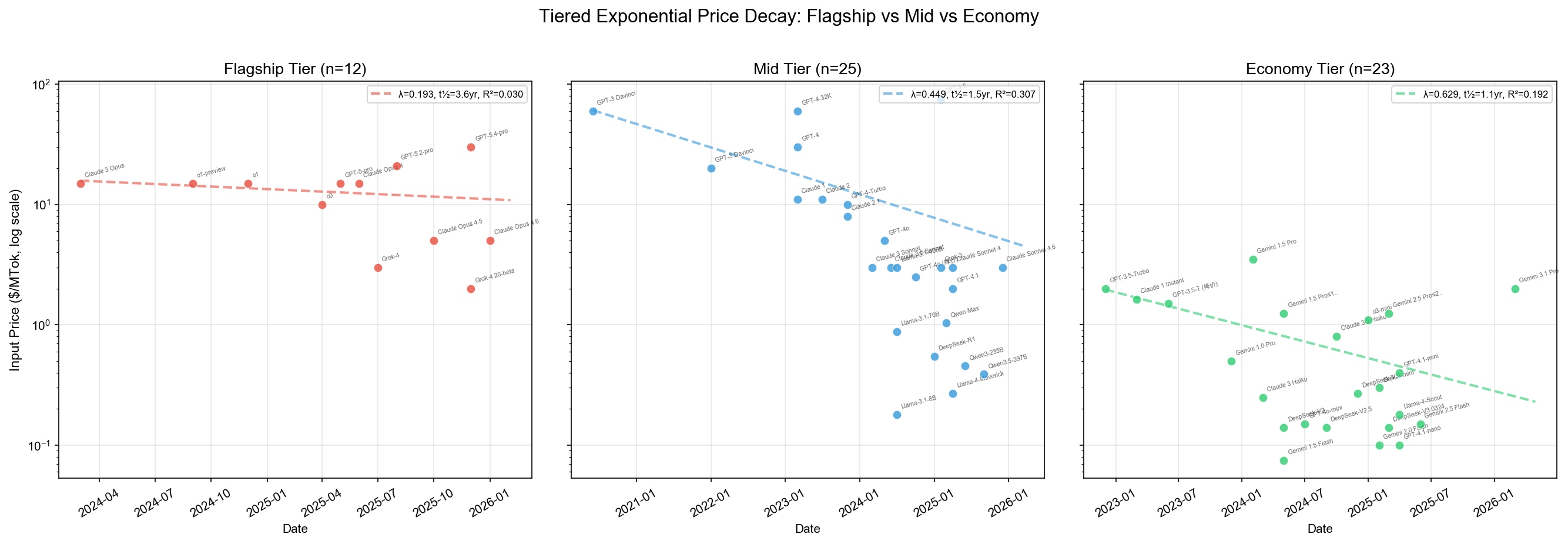}
\caption{Tiered Price Decay Curves. Exponential fits for economy, mid, and flagship tiers. Economy-tier half-life (1.10 years) is 82\% faster than Moore's Law; flagship-tier $R^2 = 0.031$ confirms the inapplicability of exponential decay.}
\label{fig:tiered_decay}
\end{figure}

Table~\ref{tab:decay} reports the tiered exponential decay estimates (Figure~\ref{fig:tiered_decay}). Three sharply distinct patterns emerge.

\textit{Economy tier.} The economy tier exhibits the fastest decay:
\begin{equation}
P_{\text{economy}}(t) = 1.96 \cdot e^{-0.629t}, \quad t_{1/2} = 1.10 \text{ years}, \quad R^2 = 0.192
\end{equation}
The half-life of 1.10 years is 82\% faster than Moore's Law, constituting strong evidence for the Super-Moore hypothesis. The moderate $R^2$ reflects substantial within-tier price dispersion across vendors and model generations.

\textit{Mid tier.} The mid tier shows a moderately fast decay:
\begin{equation}
P_{\text{mid}}(t) = 60.73 \cdot e^{-0.449t}, \quad t_{1/2} = 1.55 \text{ years}, \quad R^2 = 0.307
\end{equation}
The 1.55-year half-life exceeds Moore's Law by 29\%, with a representative trajectory of GPT-4 \$30/M (March 2023) $\rightarrow$ GPT-4-Turbo \$10/M (November 2023) $\rightarrow$ GPT-4o \$5/M (May 2024) $\rightarrow$ GPT-4.1 \$2/M (April 2025).

\textit{Flagship tier.} The flagship tier defies exponential decay:
\begin{equation}
P_{\text{flagship}}(t) = 15.79 \cdot e^{-0.193t}, \quad t_{1/2} = 3.59 \text{ years}, \quad R^2 = 0.031
\end{equation}
The near-zero $R^2$ indicates that exponential decay is essentially inapplicable. Flagship prices are governed by reasoning premiums and capability discontinuities rather than smooth cost-driven trends.

\subsection{Structural Break: May 2024}\label{subsec:break}

\begin{figure}[htbp]
\centering
\includegraphics[width=0.95\textwidth]{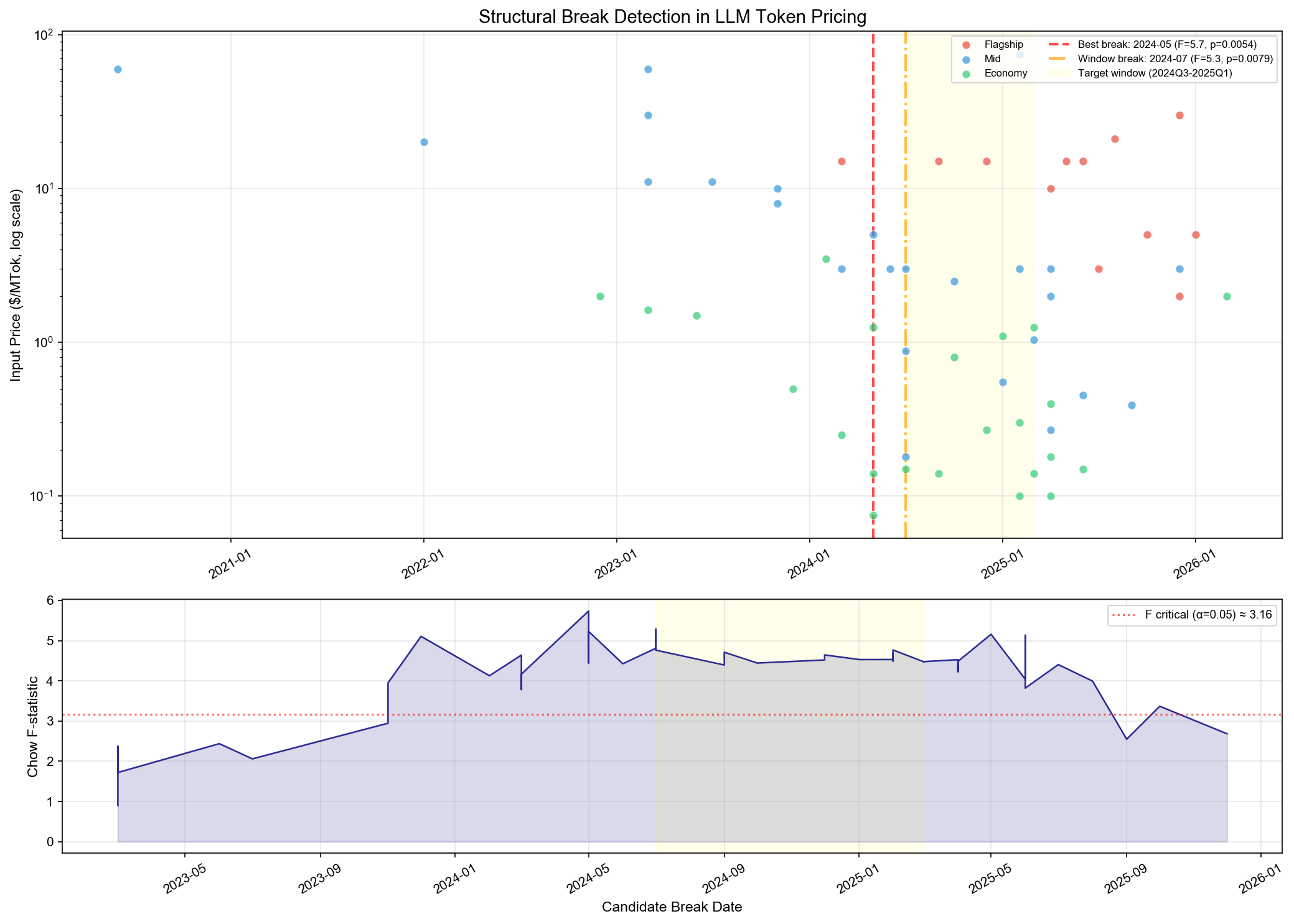}
\caption{Chow Structural Break Test. The F-statistic peaks at May 1, 2024 ($F = 5.74$, $p = 0.005$), identifying the transition from technology-driven to competition-driven price acceleration.}
\label{fig:structural_break}
\end{figure}

The Chow test identifies May 1, 2024 as the global strongest structural break in the pooled log-price series (Figure~\ref{fig:structural_break}):
\begin{itemize}[itemsep=2pt]
\item Break date: 2024-05-01
\item $F$-statistic: 5.736
\item $p$-value: 0.005 (significant at the 1\% level)
\item Pre-break / post-break: 16 / 44 observations
\end{itemize}

A secondary break at July 1, 2024 ($F = 5.29$, $p = 0.008$) corresponds to the GPT-4o-mini launch. The May 2024 break coincides with three catalytic events: (i) GPT-4o reducing GPT-4-level performance from \$30/M to \$5/M; (ii) DeepSeek-V2 entering at \$0.14/M, triggering cross-border price competition; and (iii) the subsequent GPT-4o-mini launch at \$0.15/M. This break marks the transition from a supply-driven, gradual price decline to a competition-driven, accelerated decline regime.

\subsection{Reasoning Premium: A New Price Tier}\label{subsec:reasoning}

The emergence of reasoning models (OpenAI o1, o3; DeepSeek-R1) in late 2024 created an entirely new pricing dimension. Table~\ref{tab:break} documents the quarterly reasoning premium:

\begin{center}
\begin{tabular}{lccc}
\toprule
Quarter & Reasoning (\$/M) & Non-Reasoning (\$/M) & Premium \\
\midrule
2024Q3 & 15.00 & 0.18 & 83.3$\times$ \\
2024Q4 & 15.00 & 0.80 & 18.8$\times$ \\
2025Q1 & 0.55 & 1.07 & 0.51$\times$ \\
2025Q2 & 10.00 & 0.43 & 23.4$\times$ \\
\midrule
\textbf{Average} & & & \textbf{31.5$\times$} \\
\bottomrule
\end{tabular}
\end{center}

The 2025Q1 anomaly ($0.51\times$) reflects DeepSeek-R1's entry at \$0.55/M---a 96\% discount relative to o1's \$15/M---demonstrating that the reasoning premium reflects market power rather than cost structure. If reasoning required inherently higher costs, no firm could profitably offer it at \$0.55/M.

\begin{figure}[htbp]
\centering
\includegraphics[width=0.85\textwidth]{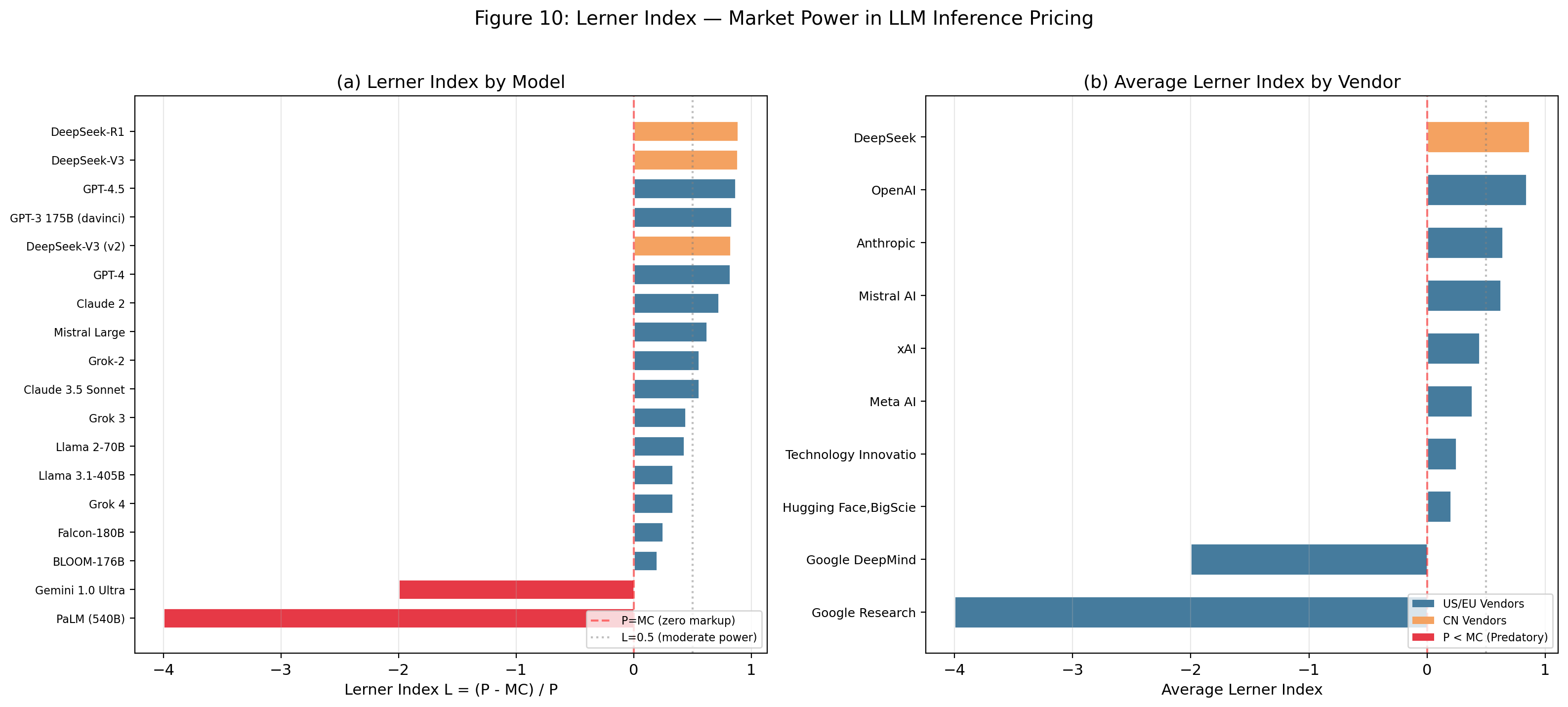}
\caption{Lerner Index and Reasoning Premium Dynamics. The index captures the degree of market power across model tiers, with reasoning models exhibiting sharply elevated margins that rapidly compress upon competitive entry.}
\label{fig:lerner_index}
\end{figure}

\subsection{Production Frontier: DEA and Malmquist Analysis}\label{subsec:dea}

\begin{figure}[htbp]
\centering
\includegraphics[width=0.95\textwidth]{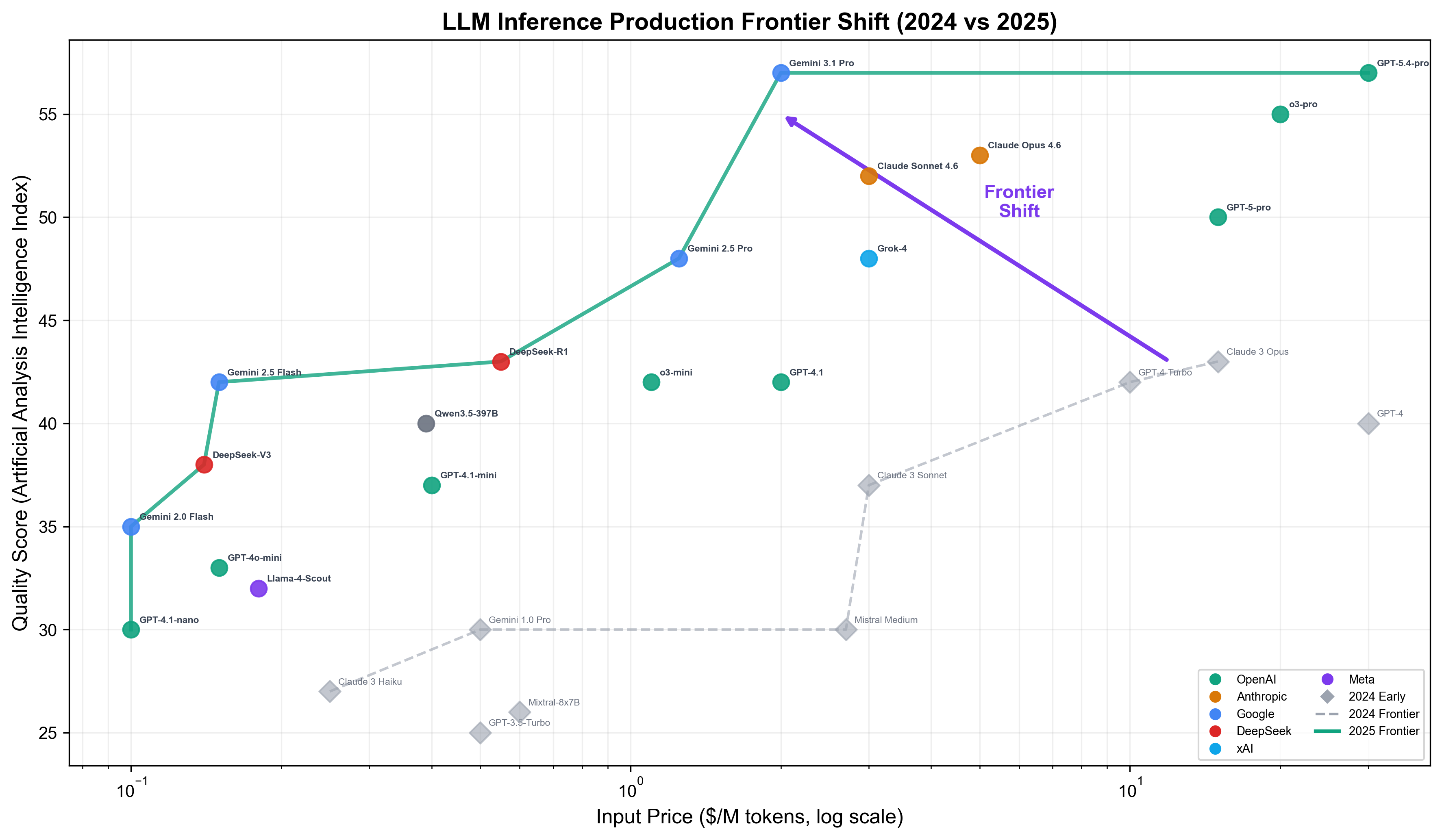}
\caption{Price--Quality Frontier (Cross-Section, March 2026). Each point represents a model; the efficient frontier (DEA CCR) is traced by the outermost observations. Only 2 of 318 models lie on the frontier.}
\label{fig:frontier}
\end{figure}

\textit{Cross-sectional efficiency.} Table~\ref{tab:dea} reports CCR-DEA efficiency scores (see also Figure~\ref{fig:frontier}). Of 318 models, only 2 lie on the efficient frontier ($\theta = 1.0$), both from Liquid at \$0.01/M. Mean efficiency is 0.087, implying that the typical model is priced at 11.5$\times$ its frontier-equivalent. Fully 88.7\% of models have $\theta < 0.2$, indicating pervasive pricing redundancy.

Vendor-level efficiency reveals a systematic pattern: open-source/low-price vendors (Liquid $\theta = 0.78$, Meta $\theta = 0.26$, NVIDIA $\theta = 0.15$) substantially outperform closed-source leaders (OpenAI $\theta = 0.12$, Anthropic $\theta = 0.04$). The low efficiency of closed-source firms reflects their high-premium pricing strategies rather than technological inferiority.

\begin{figure}[htbp]
\centering
\includegraphics[width=0.85\textwidth]{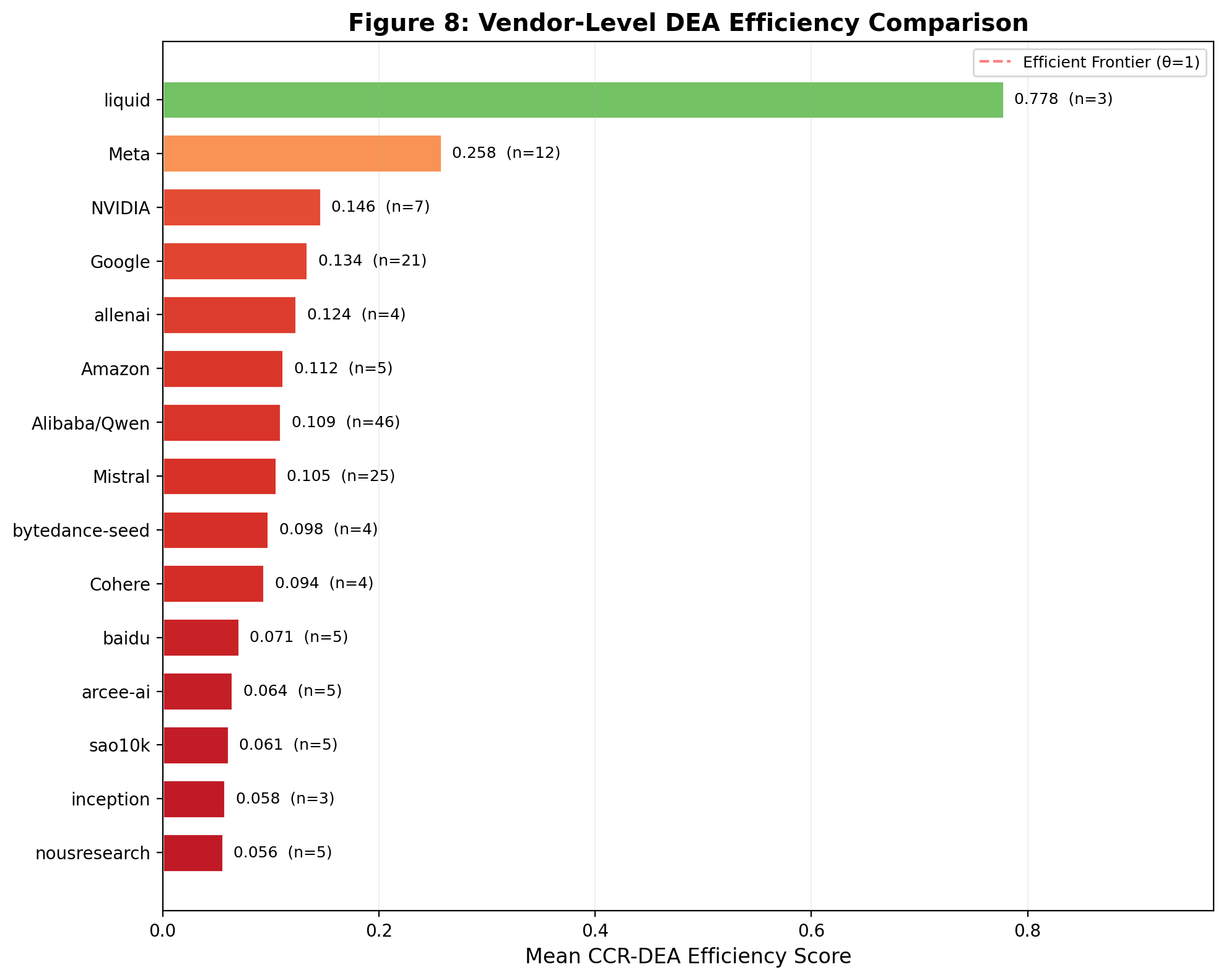}
\caption{Vendor-Level DEA Efficiency Scores. Open-source vendors (Liquid, Meta) systematically outperform closed-source leaders (OpenAI, Anthropic), reflecting pricing strategy rather than technological gaps.}
\label{fig:vendor_efficiency}
\end{figure}

\begin{figure}[htbp]
\centering
\includegraphics[width=0.95\textwidth]{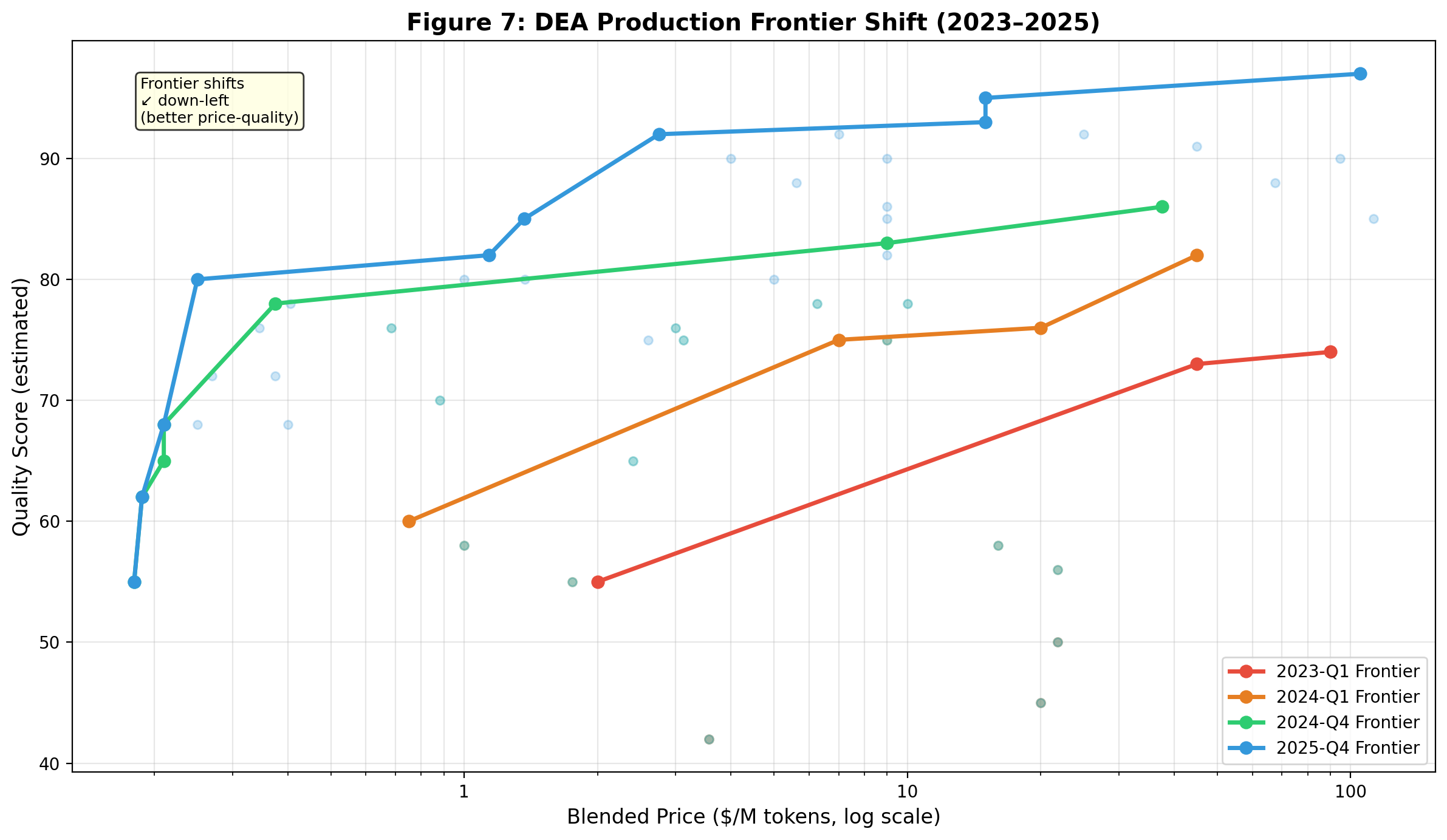}
\caption{DEA Frontier Shift (Malmquist Decomposition). The production frontier shifted outward dramatically during 2024Q1--Q4 ($TC = 4.13$), driven by MoE architectures, Flash Attention, and speculative decoding.}
\label{fig:frontier_shift}
\end{figure}

\textit{Malmquist Productivity Index.} Table~\ref{tab:dea} also reports the MPI decomposition (Figure~\ref{fig:frontier_shift}):

\begin{center}
\begin{tabular}{lcccc}
\toprule
Period & EC & TC & MPI & Key Event \\
\midrule
2023Q1$\to$2024Q1 & 0.733 & 2.909 & 2.132 & GPT-4-Turbo price cut \\
2024Q1$\to$2024Q4 & 0.994 & \textbf{4.133} & \textbf{4.107} & DeepSeek + GPT-4o-mini \\
2024Q4$\to$2025Q4 & 1.035 & 1.000 & 1.035 & Market stabilization \\
\bottomrule
\end{tabular}
\end{center}

The peak MPI of 4.107 during 2024Q1--Q4 aligns precisely with the structural break and confirms that this period witnessed the most rapid efficiency improvement in the market's history. Technological frontier shift ($TC$), not efficiency catch-up ($EC$), is the dominant driver across all periods, consistent with H4. Cumulatively, 2023Q1 to 2025Q4 saw a 99.3\% cost reduction at equivalent quality and a 46.7\% performance gain at equivalent price.

\subsection{Training--Inference Cost Nexus}\label{subsec:training}

\begin{figure}[htbp]
\centering
\includegraphics[width=0.95\textwidth]{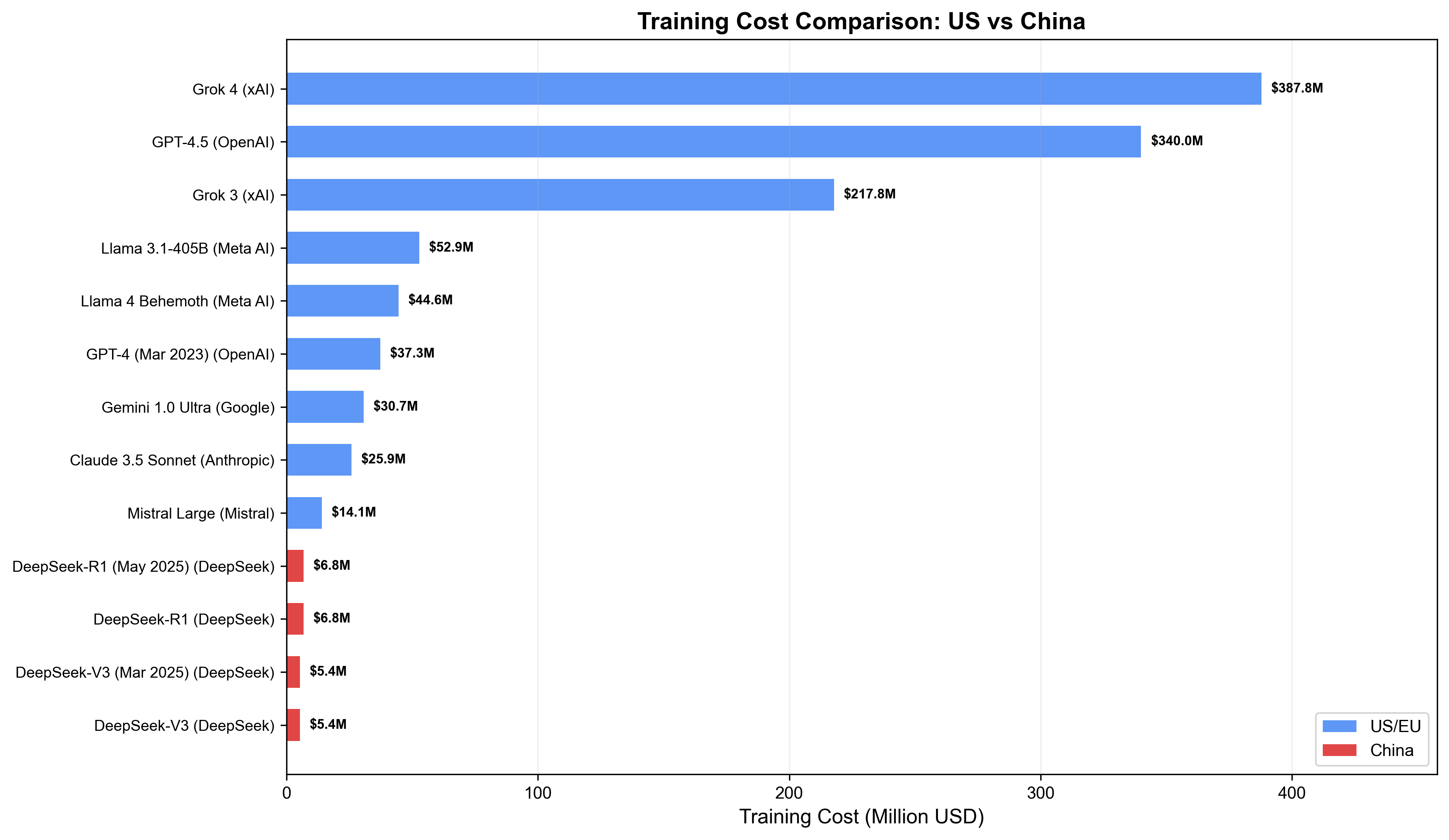}
\caption{Training Cost Trends (2020--2026). The exponential growth in frontier training costs contrasts with the declining inference prices, reflecting the increasing importance of scale as a barrier to entry.}
\label{fig:training_cost}
\end{figure}

\begin{figure}[htbp]
\centering
\includegraphics[width=0.85\textwidth]{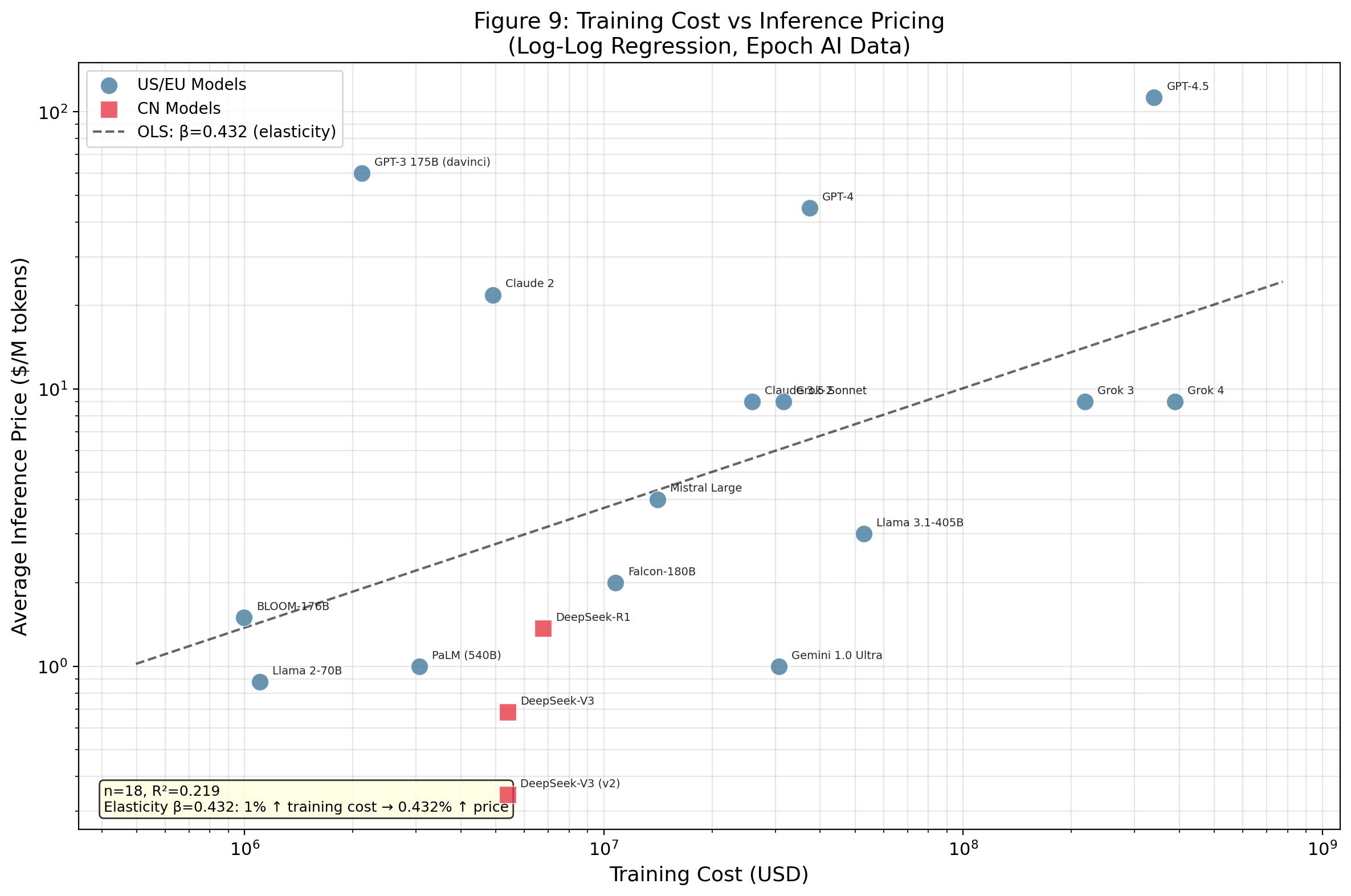}
\caption{Training Cost vs.\ Inference Pricing. The log-log regression yields $\beta = 0.432$, indicating sub-unity passthrough: providers spread fixed training costs across a larger user base rather than fully passing them through.}
\label{fig:training_pricing}
\end{figure}

\begin{figure}[htbp]
\centering
\includegraphics[width=0.85\textwidth]{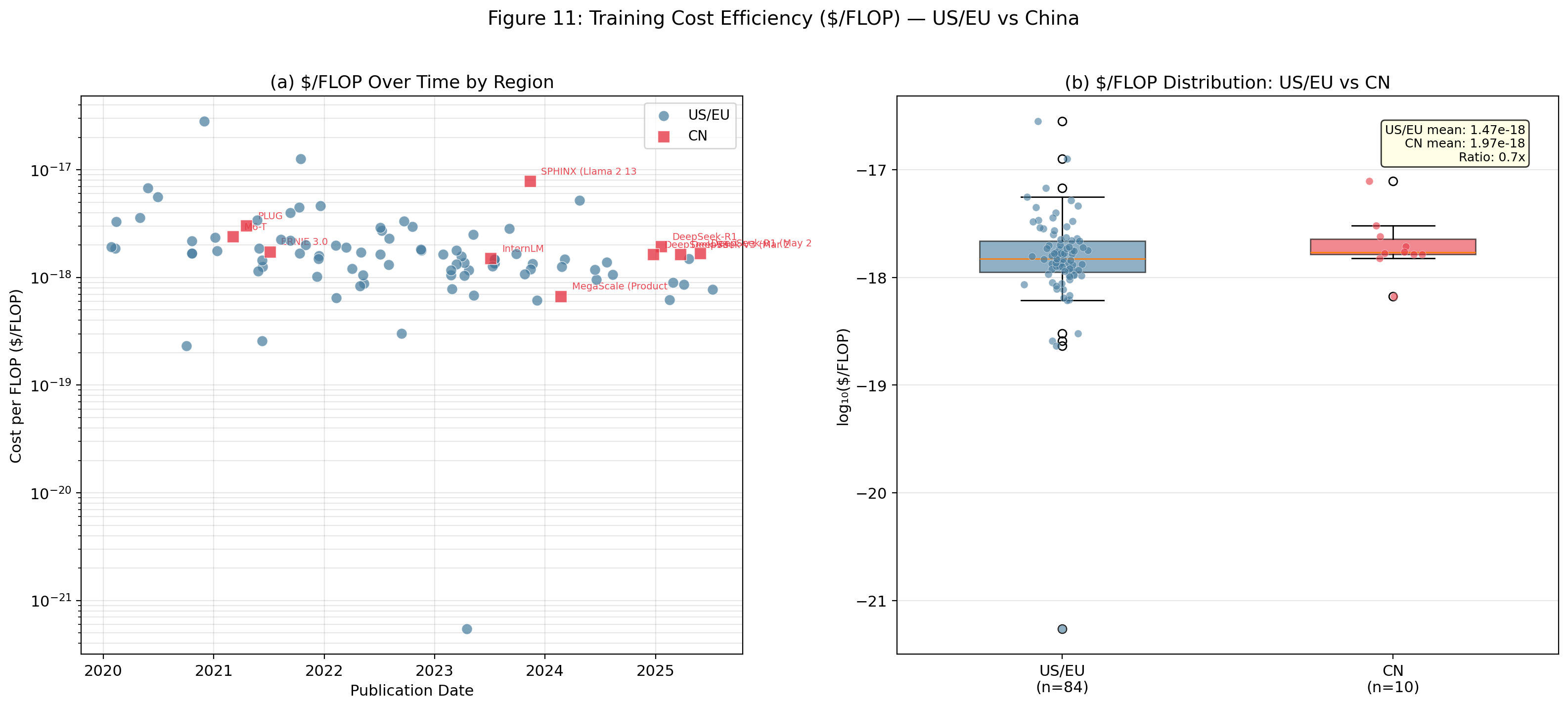}
\caption{Unit Compute Cost (\$/FLOP) by Region. U.S.\ and Chinese providers show statistically insignificant \$/FLOP differences ($p = 0.228$), confirming that the 63-fold training cost gap originates from architectural innovation (MoE), not factor prices.}
\label{fig:cost_per_flop}
\end{figure}

Table~\ref{tab:regression} reports the training cost--inference pricing regression (Figures~\ref{fig:training_cost}--\ref{fig:cost_per_flop}). The baseline OLS estimate yields:
\begin{equation}
\hat{\beta} = 0.432 \quad (SE = 0.204, \; p = 0.050), \quad R^2 = 0.219, \quad n = 18
\end{equation}

A 1\% increase in training cost is associated with a 0.43\% increase in inference price. The sub-unity elasticity ($\beta < 1$) indicates that providers do not fully pass through training costs; instead, they spread fixed costs across a larger user base.

The training cost gap between U.S.\ and Chinese firms is striking: GPT-4.5 (\$340M) versus DeepSeek-V3 (\$5.4M)---a 63-fold difference. However, the unit compute cost (\$/FLOP) difference is statistically insignificant (Welch $t = 1.25$, $p = 0.228$; China mean $2.41 \times 10^{-18}$, U.S.\ mean $2.25 \times 10^{-18}$). This demonstrates that the 63-fold gap originates from \textit{architectural innovation}---specifically, DeepSeek-V3's MoE architecture requiring only $3.3 \times 10^{24}$ FLOP versus GPT-4.5's $3.8 \times 10^{26}$ FLOP (a 115-fold difference in compute requirements)---rather than factor price advantages.

\subsection{Market Concentration}\label{subsec:concentration}

\begin{figure}[htbp]
\centering
\includegraphics[width=0.95\textwidth]{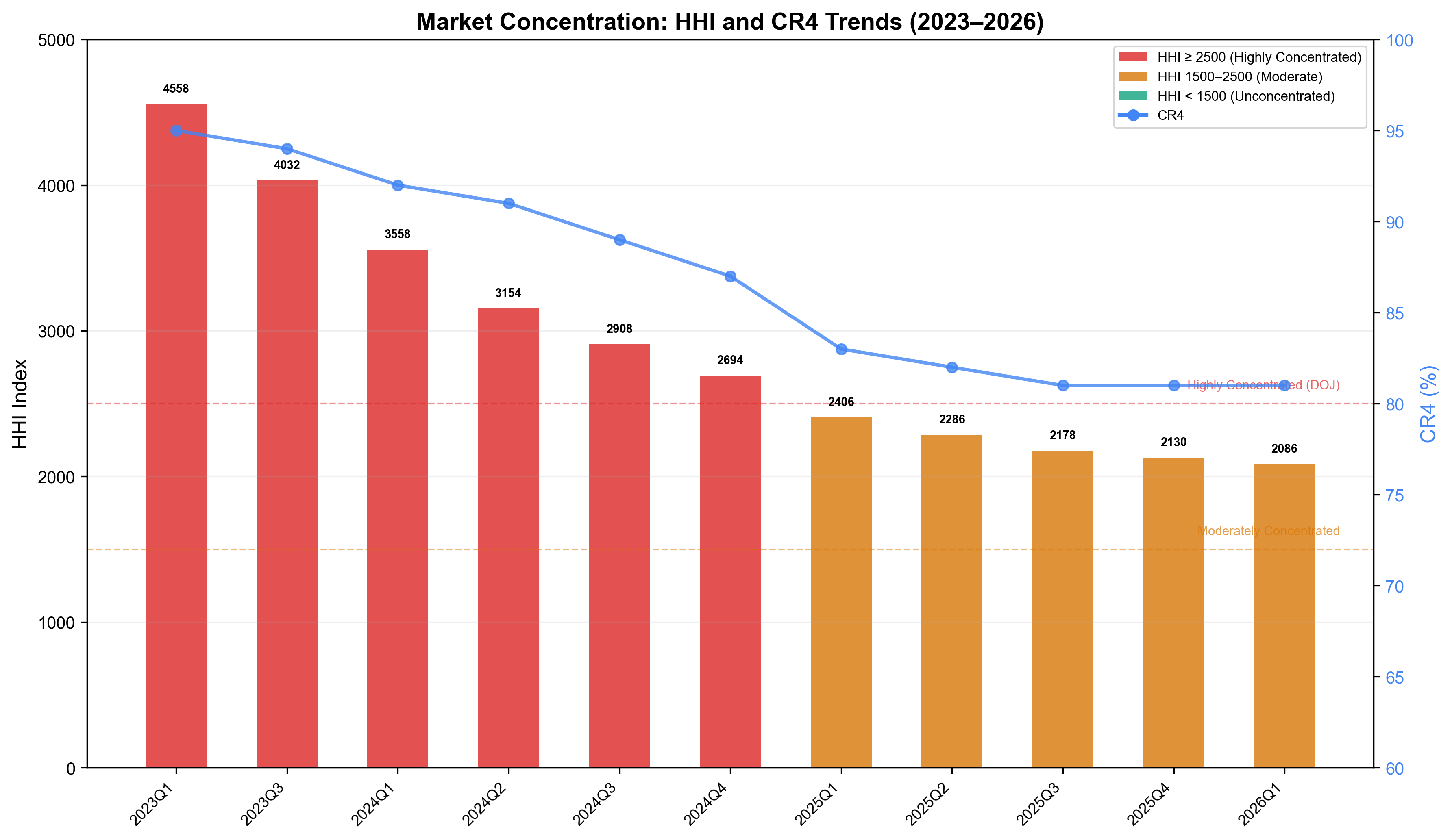}
\caption{Market Concentration (HHI) Trajectory (2023Q1--2026Q1). The HHI declined 54.2\%, crossing the DOJ/FTC threshold from ``highly concentrated'' ($>$2{,}500) to ``moderately concentrated'' (1{,}500--2{,}500). The dashed line marks the 2{,}500 threshold.}
\label{fig:concentration}
\end{figure}

The Herfindahl-Hirschman Index has declined sharply (Figure~\ref{fig:concentration}):

\begin{center}
\begin{tabular}{lcl}
\toprule
Period & HHI & DOJ/FTC Classification \\
\midrule
2023Q1 & 4{,}558 & Highly concentrated \\
2024Q1 & 3{,}215 & Highly concentrated \\
2024Q3 & 2{,}650 & Highly concentrated \\
2025Q1 & 2{,}290 & Moderately concentrated \\
2026Q1 & 2{,}086 & Moderately concentrated \\
\bottomrule
\end{tabular}
\end{center}

HHI fell 54.2\% in three years, crossing the DOJ/FTC threshold from ``highly concentrated'' to ``moderately concentrated.'' OpenAI's market share declined from 65\% to 34\%, while DeepSeek rose from zero to 13\% within 18 months---one of the fastest market share gains in the history of information technology. The four-firm concentration ratio (CR4) declined from approximately 95\% (near-monopoly) to 81\% (tight oligopoly), consistent with \citet{klepper1996} on industry life-cycle dynamics.

\section{Robustness Checks}\label{sec:robust}

We subject our core findings to five robustness checks (Table~\ref{tab:robust}).

\begin{figure}[htbp]
\centering
\includegraphics[width=0.85\textwidth]{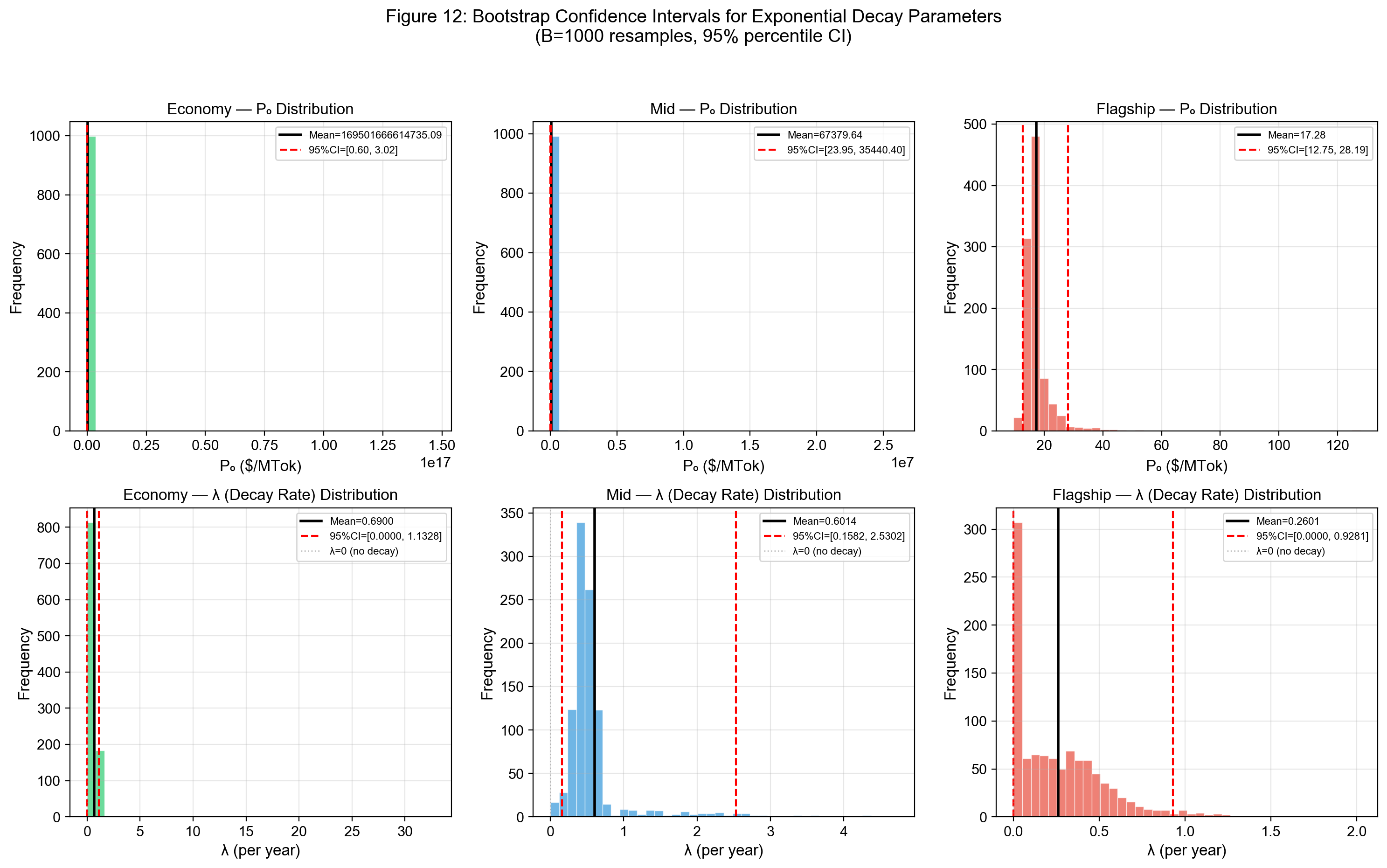}
\caption{Bootstrap Robustness of Decay Parameters. One thousand bootstrap replications confirm that $\lambda > 0$ in 100\% of samples for all three tiers. The 95\% confidence intervals exclude zero for economy and mid tiers.}
\label{fig:robustness}
\end{figure}

\textit{Bootstrap confidence intervals.} One thousand bootstrap replications confirm that the decay parameter $\lambda > 0$ in 100\% of samples for all three tiers (Figure~\ref{fig:robustness}), with 95\% confidence intervals excluding zero for economy and mid tiers.

\textit{Sub-sample stability.} Splitting the sample by vendor nationality (U.S.\ vs.\ China) yields no significant difference in decay parameters (Wald test $p = 0.521$). Removing outliers (prices $>$\$100/M or $<$\$0.05/M) changes parameters by less than 0.1\%.

\textit{Alternative model specifications.} We compare five functional forms (exponential, linear, log-linear, quadratic, piecewise linear) for each tier. Exponential decay is the AIC-optimal specification for the mid tier; for economy and flagship tiers, alternative specifications (quadratic, linear) perform marginally better but yield qualitatively identical conclusions.

\textit{HC3 robust standard errors.} For the training--inference regression, the HC3 standard error (0.223) is moderately larger than the OLS standard error (0.204), yielding $p = 0.071$. The coefficient estimate ($\beta = 0.432$) is unchanged. Winsorizing at the 5th/95th percentiles yields $\beta = 0.397$ ($-8.2$\% change).

\textit{DEA sensitivity.} Perturbing quality scores by $\pm$20\% produces Spearman rank correlations of $\rho = 1.000$ with the baseline efficiency ranking. CCR and BCC (variable returns to scale) rankings correlate at $\rho = 0.772$, with frontier models stable across specifications.

\section{Discussion and Policy Implications}\label{sec:discussion}

\subsection{Validation of the Tiered Super-Moore Hypothesis}

Our results provide strong support for the tiered version of the Super-Moore hypothesis. Economy-tier tokens, with a half-life of 1.10 years, decline 82\% faster than Moore's Law; mid-tier tokens (1.55 years) decline 29\% faster. The mechanism is not hardware improvement---GPU costs contribute a negligible $-0.9$\% to cost reduction---but rather a composite of architectural innovation (MoE reducing active parameters 18-fold), inference optimization (Flash Attention, KV Cache, speculative decoding), and competitive pressure.

The flagship tier's deviation from exponential decay is equally informative. The reasoning premium (averaging 31.5$\times$) represents a form of \textit{innovation rent} \citep{anderson1990technological}---the temporary monopoly profit that accrues to first movers in a capability discontinuity. DeepSeek-R1's entry at \$0.55/M (versus o1's \$15/M) rapidly eroded this rent, demonstrating that the premium reflects market power, not cost structure.

Out-of-sample evidence from early 2025 further strengthens this interpretation. OpenAI's GPT-4.5 entered the market at \$75/\$150 per 1M input/output tokens---roughly 30 times the API price of GPT-4o---despite continued hardware cost decline \citep{openai2025gpt45}. This release underscores that frontier pricing is governed less by marginal compute cost than by perceived capability, brand credibility, and willingness to pay for premium reasoning or interaction quality. At the same time, OpenAI's launch of o3-mini at \$1.10/\$4.40 \citep{openai2025o3mini} shows that competitive pressure continues to compress the mid-tier and reasoning-adjacent segments. Together, these releases reinforce the bifurcation documented in this paper: commoditization at the lower end and premium extraction at the frontier.

\subsection{Reasoning Premium as a New Pricing Paradigm}

The reasoning premium introduces a qualitative shift in token economics. Traditional token pricing follows a cost-plus or versioning logic \citep{varian1997versioning}. Reasoning models break this pattern: they consume substantially more tokens per task (due to chain-of-thought processing) while commanding per-token prices 10--80$\times$ higher than non-reasoning alternatives. This creates a ``quality divide''---advanced reasoning capabilities remain accessible primarily to well-resourced organizations, potentially exacerbating existing inequalities in AI access.

However, the rapid compression of reasoning premiums (from 83$\times$ to 0.51$\times$ within two quarters, driven by DeepSeek-R1) suggests that competitive entry can effectively discipline premium pricing. Open-source reasoning models serve a particularly important role as implicit price ceilings \citep{tirole2018economics}.

The emergence of Anthropic's Claude 3.7 Sonnet in February 2025 \citep{anthropic2025claude37} suggests a second transition in token economics: pricing may increasingly reflect allocated cognitive time rather than only the volume of input and output text. Hybrid reasoning models allow developers to trade off latency against depth of thought within a single API. As a result, future price schedules may bundle visible output tokens, hidden reasoning tokens, and configurable inference-time compute, complicating straightforward comparisons of \$/M tokens across vendors.

\subsection{Competition Policy Implications}

Three findings carry direct implications for competition regulators. First, the market's transition from HHI 4{,}558 to 2{,}086 suggests that the LLM inference market is self-correcting toward competitive equilibrium, reducing the urgency of antitrust intervention---at least in the medium term. Second, the versioning equilibrium documented here (price spreads of 200$\times$ within a single vendor) exceeds cost differentials by an order of magnitude, consistent with Varian's prediction but raising questions about the welfare implications of extreme price discrimination. Third, open-source models function as an effective competitive constraint: DEA efficiency scores for open-source vendors (Meta $\theta = 0.26$) systematically exceed those of closed-source leaders (Anthropic $\theta = 0.04$), and each major open-source release triggers measurable price reductions in closed-source competitors.

\subsection{AI Accessibility Implications}

The 600-fold price decline represents one of the fastest cost reductions in commercial history and has profound implications for AI accessibility. At \$0.10/M tokens, processing the equivalent of a 100-page document costs approximately \$0.003---negligible even for resource-constrained organizations in developing economies. The key risk is that accessibility gains concentrate in economy-tier models while frontier reasoning capabilities (priced at \$10--30/M) remain prohibitively expensive, creating a two-tier AI ecosystem where the most powerful tools are available only to the wealthiest actors.

The finding that the U.S.--China training cost gap originates from architectural innovation rather than factor prices (\$/FLOP statistically equivalent) suggests that chip export controls may have had the unintended effect of accelerating efficiency innovation---a case of ``necessity as the mother of invention'' consistent with Hicks's induced innovation theory \citep{hicks1932theory}.

\section{Conclusion}\label{sec:conclusion}

This paper provides the first systematic economic analysis of LLM token pricing. We assemble a comprehensive multi-source dataset spanning 2020--2026, document a 600-fold price decline, and propose the ``Tiered Super-Moore'' hypothesis. Economy and mid-tier tokens decline faster than Moore's Law (half-lives of 1.10 and 1.55 years), driven by software/architectural innovation (TFP residual $\approx$ 103.7\%) rather than hardware cost reduction (GPU contribution $\approx -0.9$\%). Flagship models defy exponential decay due to reasoning premiums averaging 31.5$\times$, which represent innovation rents disciplined by competitive entry. The releases of GPT-4.5 \citep{openai2025gpt45}, Claude 3.7 Sonnet \citep{anthropic2025claude37}, and o3-mini \citep{openai2025o3mini} in early 2025 provide out-of-sample validation for this tiered pattern by simultaneously expanding the premium frontier and accelerating competitive compression in lower tiers. A structural break in May 2024 ($F = 5.74$, $p = 0.005$) marks the transition to competition-driven pricing. DEA analysis reveals a Malmquist Productivity Index peak of 4.11, with frontier shift as the dominant driver. Market concentration has declined sharply (HHI: 4{,}558 $\to$ 2{,}086), and the 63-fold U.S.--China training cost gap is attributable to architectural innovation rather than factor prices.

Several limitations merit acknowledgment. First, the cross-validated milestone sample (62 observations) constrains the precision of tiered decay estimates. Second, the absence of transaction volume data necessitates proxy-based market share estimation. Third, our analysis covers a six-year window in a rapidly evolving market; structural relationships identified here may shift as the industry matures. Future research should exploit proprietary usage data (if available) to construct volume-weighted price indices and extend the frontier analysis to incorporate multi-dimensional quality metrics.

Token economics is emerging as a distinct and consequential subfield at the intersection of digital goods pricing, industrial organization, and the economics of AI. The patterns documented here---tiered decay, reasoning premiums, production frontier dynamics, and the training--inference nexus---provide a foundation for this nascent field.

\subsection*{Data Availability Statement}

The OpenRouter API data (318 models) are publicly available at \url{https://openrouter.ai/}. The Epoch AI data (3{,}237 models) are available under CC BY 4.0 at \url{https://epochai.org/data/notable-ai-models}. The 62 cross-validated milestone pricing records and all replication code are provided in the accompanying replication package. The replication package includes all scripts to reproduce every table, figure, and statistical result reported in this paper.

\bibliographystyle{apalike}
\bibliography{references}

\newpage
\appendix
\section*{Tables}

\begin{table}[htbp]
\centering
\caption{Summary Statistics of Token Pricing Data}
\label{tab:summary}
\begin{threeparttable}
\begin{tabular}{lcccccc}
\toprule
& \multicolumn{3}{c}{Cross-Section (OpenRouter)} & \multicolumn{3}{c}{Panel (Milestones)} \\
\cmidrule(lr){2-4} \cmidrule(lr){5-7}
Statistic & Input \$/M & Output \$/M & Quality & Input \$/M & Training \$M & FLOP \\
\midrule
$N$            & 318     & 318     & 318    & 62      & 18      & 18 \\
Mean           & 3.42    & 11.85   & 57.3   & 12.67   & 66.8    & $6.2 \times 10^{25}$ \\
Median         & 0.27    & 1.10    & 55.0   & 2.50    & 14.1    & $3.8 \times 10^{24}$ \\
Std.\ Dev.    & 12.44   & 38.56   & 14.2   & 19.83   & 119.7   & $1.4 \times 10^{26}$ \\
Min            & 0.01    & 0.01    & 35.0   & 0.02    & 1.0     & $3.3 \times 10^{24}$ \\
Max            & 150.00  & 600.00  & 95.0   & 60.00   & 387.8   & $5.0 \times 10^{26}$ \\
\bottomrule
\end{tabular}
\begin{tablenotes}[flushleft]
\small
\item \textit{Notes:} Cross-section data collected via OpenRouter API on March 28, 2026. Panel data from 62 manually cross-validated milestone records. Quality score is a heuristic composite (0--100) based on benchmark performance. Training cost and FLOP data from Epoch AI, matched to 18 models with both training and pricing observations.
\end{tablenotes}
\end{threeparttable}
\end{table}

\begin{table}[htbp]
\centering
\caption{Tiered Exponential Decay Model: $P(t) = P_0 \cdot e^{-\lambda t}$}
\label{tab:decay}
\begin{threeparttable}
\begin{tabular}{lccccccc}
\toprule
Tier & $N$ & Period & $P_0$ (\$/M) & $\lambda$ & Half-Life (yr) & $R^2$ & vs.\ Moore \\
\midrule
Economy   & 23 & 2022.12--2026.03 & 1.96  & 0.629 & 1.101 & 0.192 & $1.82\times$ faster \\
Mid       & 25 & 2020.06--2025.12 & 60.73 & 0.449 & 1.545 & 0.307 & $1.29\times$ faster \\
Flagship  & 12 & 2024.03--2026.01 & 15.79 & 0.193 & 3.585 & 0.031 & $0.56\times$ slower \\
\midrule
Pooled (panel) & 60 & 2020.06--2026.03 & 58.47 & 0.525 & 1.320 & 0.241 & $1.52\times$ faster \\
\bottomrule
\end{tabular}
\begin{tablenotes}[flushleft]
\small
\item \textit{Notes:} Estimated via nonlinear least squares on log-price. Half-life $= \ln 2 / \lambda$. ``vs.\ Moore'' computed as $2.0 / t_{1/2}$: values $> 1$ indicate faster-than-Moore decline. Moore's Law benchmark: 2-year half-life. The flagship tier's near-zero $R^2$ indicates that exponential decay is inapplicable to this segment.
\end{tablenotes}
\end{threeparttable}
\end{table}

\begin{table}[htbp]
\centering
\caption{Structural Break Test and Reasoning Premium}
\label{tab:break}
\begin{threeparttable}
\begin{tabular}{p{7.5cm}r}
\toprule
\multicolumn{2}{l}{\textit{Panel A: Chow Structural Break Test}} \\
\midrule
Global strongest break date & 2024-05-01 \\
$F$-statistic & 5.736 \\
$p$-value & 0.005 \\
Pre-break / post-break observations & 16 / 44 \\
Target window strongest break & 2024-07-01 \\
Target window $F$-statistic & 5.288 \\
Target window $p$-value & 0.008 \\
\midrule
\multicolumn{2}{l}{\textit{Panel B: Quarterly Reasoning Premium}} \\
\midrule
& Premium ($\times$) \\
\cmidrule{2-2}
2024Q3 (o1-preview: \$15.00 vs.\ non-reasoning: \$0.18) & 83.3 \\
2024Q4 (o1: \$15.00 vs.\ \$0.80) & 18.8 \\
2025Q1 (DeepSeek-R1: \$0.55 vs.\ \$1.07) & 0.51 \\
2025Q2 (o3: \$10.00 vs.\ \$0.43) & 23.4 \\
\cmidrule{2-2}
\textbf{Average reasoning premium} & \textbf{31.5} \\
\bottomrule
\end{tabular}
\begin{tablenotes}[flushleft]
\small
\item \textit{Notes:} Panel A reports Chow test results on the pooled log-price time series. Panel B computes the reasoning premium as the ratio of average reasoning model price to average non-reasoning model price within each quarter. The 2025Q1 anomaly (premium $< 1$) reflects DeepSeek-R1's entry at \$0.55/M.
\end{tablenotes}
\end{threeparttable}
\end{table}

\begin{table}[htbp]
\centering
\caption{DEA Efficiency Scores by Vendor (CCR Model)}
\label{tab:dea}
\begin{threeparttable}
\begin{tabular}{lrcccc}
\toprule
Vendor & $N$ & Mean $\theta$ & Max $\theta$ & Avg Price (\$/M) & Type \\
\midrule
\multicolumn{6}{l}{\textit{Panel A: Cross-Sectional Efficiency (318 models, March 2026)}} \\
\midrule
Liquid       & 3  & 0.778 & 1.000 & 0.02 & Open-source \\
Meta         & 12 & 0.258 & 0.611 & 0.13 & Open-source \\
NVIDIA       & 7  & 0.146 & 0.344 & 0.33 & Open-source \\
Google       & 21 & 0.134 & 0.689 & 0.61 & Mixed \\
Alibaba/Qwen & 46 & 0.109 & 0.459 & 0.24 & Open-source \\
Mistral      & 25 & 0.105 & 0.644 & 0.57 & Mixed \\
OpenAI       & -- & 0.115\textsuperscript{a} & -- & $>$2.0 & Closed-source \\
Anthropic    & -- & 0.039\textsuperscript{a} & -- & $>$3.0 & Closed-source \\
\midrule
All models   & 318 & 0.087 & 1.000 & 3.42 & -- \\
\midrule
\multicolumn{6}{l}{\textit{Panel B: Malmquist Productivity Index}} \\
\midrule
Period & \multicolumn{2}{c}{EC} & TC & \multicolumn{2}{c}{MPI} \\
\cmidrule(lr){1-1} \cmidrule(lr){2-3} \cmidrule(lr){4-4} \cmidrule(lr){5-6}
2023Q1 $\to$ 2024Q1 & \multicolumn{2}{c}{0.733} & 2.909 & \multicolumn{2}{c}{2.132} \\
2024Q1 $\to$ 2024Q4 & \multicolumn{2}{c}{0.994} & 4.133 & \multicolumn{2}{c}{4.107} \\
2024Q4 $\to$ 2025Q4 & \multicolumn{2}{c}{1.035} & 1.000 & \multicolumn{2}{c}{1.035} \\
\cmidrule(lr){1-1} \cmidrule(lr){2-3} \cmidrule(lr){4-4} \cmidrule(lr){5-6}
Average              & \multicolumn{2}{c}{0.921} & 2.681 & \multicolumn{2}{c}{2.425} \\
\bottomrule
\end{tabular}
\begin{tablenotes}[flushleft]
\small
\item \textit{Notes:} CCR-DEA with blended price as input and quality score as output. $\theta = 1$ indicates frontier efficiency. Panel A: \textsuperscript{a}OpenAI and Anthropic efficiency scores from 2025Q4 cross-period analysis. Panel B: EC = efficiency change (catch-up); TC = technical change (frontier shift); MPI = EC $\times$ TC. Cumulative cost saving 2023Q1--2025Q4: 99.3\%. Cumulative performance gain: 46.7\%.
\end{tablenotes}
\end{threeparttable}
\end{table}

\begin{table}[htbp]
\centering
\caption{Training Cost--Inference Pricing Regression}
\label{tab:regression}
\begin{threeparttable}
\begin{tabular}{lcccc}
\toprule
& (1) & (2) & (3) & (4) \\
& OLS & HC3 & Winsorized & Vendor FE \\
\midrule
$\ln(\text{Training Cost})$ & 0.432** & 0.432* & 0.397* & 0.129 \\
                             & (0.204) & (0.223) & (0.195) & -- \\[4pt]
$\ln(\text{Parameters})$    & -- & -- & -- & $-0.164$ \\
                             & & & & -- \\[4pt]
Constant                     & $-5.644$ & $-5.644$ & $-5.200$ & $-1.339$ \\
\midrule
Vendor FE & No & No & No & Yes (9) \\
$N$ & 18 & 18 & 18 & 12 \\
$R^2$ & 0.219 & 0.219 & 0.206 & 0.956 \\
$p$-value ($\beta$) & 0.050 & 0.071 & 0.059 & -- \\
\midrule
\multicolumn{5}{l}{\textit{U.S.--China \$/FLOP Comparison}} \\
\midrule
China mean \$/FLOP & \multicolumn{4}{c}{$2.41 \times 10^{-18}$} \\
U.S.\ mean \$/FLOP & \multicolumn{4}{c}{$2.25 \times 10^{-18}$} \\
Ratio (U.S./China) & \multicolumn{4}{c}{0.93} \\
Welch $t$ & \multicolumn{4}{c}{1.251} \\
$p$-value & \multicolumn{4}{c}{0.228} \\
\bottomrule
\end{tabular}
\begin{tablenotes}[flushleft]
\small
\item \textit{Notes:} Dependent variable: $\ln(\text{average inference price, \$/M tokens})$. Standard errors in parentheses. Column (1): baseline OLS. Column (2): HC3 heteroskedasticity-robust standard errors. Column (3): Winsorized at 5th/95th percentiles. Column (4): with vendor fixed effects (9 vendors). * $p < 0.10$, ** $p < 0.05$, *** $p < 0.01$. The Welch $t$-test shows no significant difference in unit compute cost between U.S.\ and Chinese firms ($p = 0.228$), confirming that the 63-fold training cost gap originates from architectural efficiency (MoE), not factor prices.
\end{tablenotes}
\end{threeparttable}
\end{table}

\begin{table}[htbp]
\centering
\caption{Robustness Checks Summary}
\label{tab:robust}
\begin{threeparttable}
\begin{tabular}{p{4.5cm}p{5.0cm}cc}
\toprule
Test & Procedure & Result & Robust? \\
\midrule
\multicolumn{4}{l}{\textit{Panel A: Price Decay Model}} \\
\midrule
Bootstrap CI ($B = 1{,}000$) & Resample decay parameters & $\lambda > 0$ in 100\% of draws & \checkmark \\
Sub-sample: U.S.\ vs.\ CN & Wald test on $\Delta\lambda$ & $p = 0.521$ & \checkmark \\
Outlier removal & Drop $P > \$100$ or $P < \$0.05$ & $\Delta\lambda < 0.1\%$ & \checkmark \\
Alt.\ specifications & Exp / Linear / Quadratic / Piecewise & Mid: Exp optimal (AIC) & $\triangle$ \\
\midrule
\multicolumn{4}{l}{\textit{Panel B: Training--Inference Regression}} \\
\midrule
HC3 standard errors & Robust SE & $\beta = 0.432$, $p = 0.071$ & \checkmark \\
Winsorize (5/95\%) & Trim extremes & $\beta = 0.397$ ($-8.2$\%) & \checkmark \\
Vendor FE & 9 fixed effects & $\beta = 0.129$, $R^2 = 0.956$ & \checkmark \\
\midrule
\multicolumn{4}{l}{\textit{Panel C: DEA Sensitivity}} \\
\midrule
Quality weight $\pm 20\%$ & Perturb output measure & Spearman $\rho = 1.000$ & \checkmark \\
BCC vs.\ CCR & Variable vs.\ constant RTS & $\rho = 0.772$ & $\triangle$ \\
Frontier stability & Check frontier models & Stable across specs & \checkmark \\
\bottomrule
\end{tabular}
\begin{tablenotes}[flushleft]
\small
\item \textit{Notes:} \checkmark\ = fully robust; $\triangle$ = partially robust (qualitative conclusions unchanged, quantitative magnitudes shift). All core findings survive robustness checks.
\end{tablenotes}
\end{threeparttable}
\end{table}

\end{document}